\documentclass[referee]{raa}            
\usepackage{graphicx,times}             
\usepackage{natbib}
\usepackage{amssymb,amsmath}
\usepackage{float} 
\bibpunct{(}{)}{;}{a}{}{,}
\usepackage[pagebackref=true]{hyperref}

\usepackage{multirow}

\begin{document}

  \title{JW-VL: A Vision-Language Model for Solar Physics
}

   \volnopage{Vol.0 (20xx) No.0, 000--000}      
   \setcounter{page}{1}          

   \author{Mingfu Shao 
      \inst{1,2}
   \and Hui Wang
      \inst{1,2}
   \and Liyue Tong
      \inst{1}
   \and Yuyang Li
      \inst{2,1}
   \and Cunshi Wang
      \inst{2,1}
   \and Jiaben Lin \footnote{Corresponding Author: jiabenlin@bao.ac.cn}
      \inst{1,2}
   \and Suo Liu
      \inst{1,2}
   \and Haiqing Xu
      \inst{1,2}
   \and Yin Zhang
      \inst{1,2}
   \and Jing Huang
      \inst{1,2}
   }

   \institute{State Key Laboratory of Solar Activity and Space Weather, NAOC, Beijing 100101, P. R. China; {\it shaomf@bao.ac.cn}
        \and
             University of Chinese Academy of Sciences, Beijing 101408, P. R. China;
\vs\no
   {\small Received 20xx month day; accepted 20xx month day}}
\abstract{ 
Vision-Language Models (VLMs) have achieved breakthrough progress in general knowledge domains, yet adaptation to specialized scientific fields remains challenging due to multimodal representation shifts and the limited integration of domain-specific knowledge. To address the limitations of general-purpose VLMs when applied to solar physics image recognition, analysis, and reasoning,  we propose JinWu Vision–Language (JW-VL), a fine-tuned foundation model tailored for solar physics. The model integrates multi-wavelength observational data from both space-based and ground-based telescopes, encompassing representative spectral bands spanning the photosphere, chromosphere, and corona. Built upon a cross-modal alignment knowledge distillation framework, JW-VL learns a joint visual–semantic embedding that enables end-to-end modeling from raw solar observational data to downstream tasks, including solar image recognition, solar activity analysis via image-based question answering, and optical character recognition (OCR), while also supporting the construction of a multi-band, cross-instrument solar image benchmark dataset. Furthermore, as a demonstration of interdisciplinary applicability, we developed a “Daily Solar Activity Reports” agent comprising core modules for solar activity level assessment, significant active region characterization, magnetic field complexity analysis, potential space weather impact assessment, and  identifying active regions for targeted observation. While JW-VL may not yet meet the rigorous, high-precision demands of operational solar physics, it bridges raw observations and diverse downstream tasks, establishing a valuable methodological framework for applying multimodal deep learning to the field. 
}

   \authorrunning{Mingfu Shao, Hui Wang, \& Liyue Tong et al.}            
   \titlerunning{JW-VL: A Vision-Language Model for Solar Physics with Applications }  

   \maketitle
%
%
\section{Introduction}           
\label{sect:intro}

\begin{figure}
   \centering
   \includegraphics[width=\textwidth, angle=0]{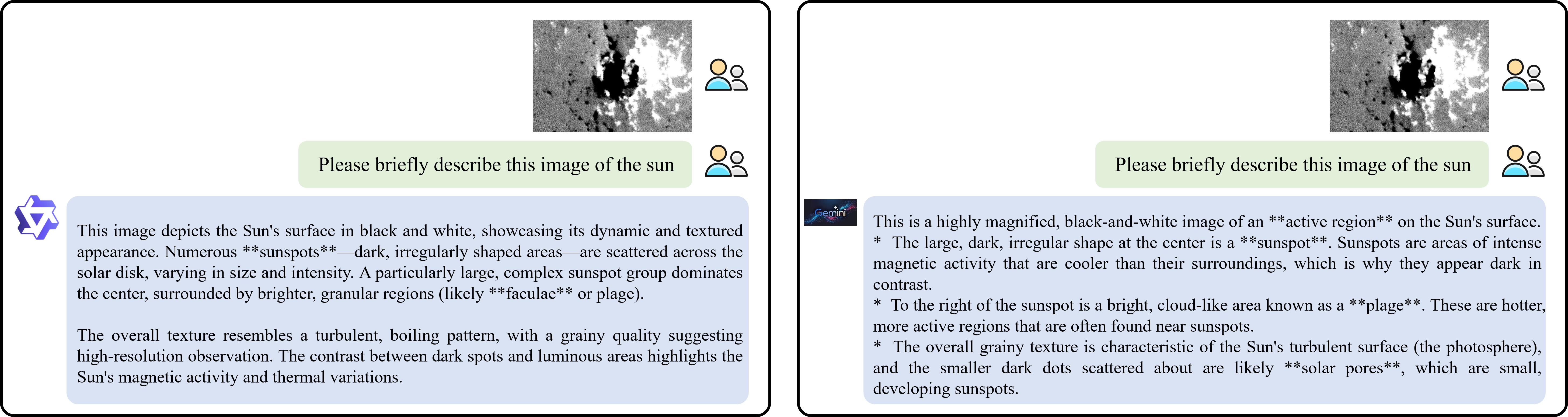}
   \caption{The limitations of the QVQ-Max and Gemini 2.5 Pro models in interpreting specialized solar magnetogram. Both responses misinterpret the image as an intensity-based photospheric observation rather than a line-of-sight magnetogram. As a result, magnetic field signals are incorrectly described in terms of brightness, temperature, sunspots, and plage, leading to physical mischaracterization of the observed structures.}
   \label{fig:general_model}
   \end{figure}

In recent years, large language models (LLMs) have attracted considerable attention due to their strong capabilities in natural language generation and comprehension. However, conventional LLMs are predominantly trained on textual data from public web sources and lack the ability to process multimodal information such as images, audio, and video. To overcome this limitation, vision–language models (VLMs) have been developed to integrate visual perception with language understanding through large-scale image–text training, with performance further enhanced by scaling model capacity and training data. VLMs have achieved remarkable success in general-purpose visual tasks, demonstrating broad applicability, and their extension to domain-specific scientific fields has facilitated progress across multiple disciplines (\citealt{dai2025improving, gao2024physicallygroundedvisionlanguagemodels, van2024largevisuallanguagemodels}). For example, in the biomedical domain, the LLaVA-Med model constructs multimodal instruction-following datasets and enables end-to-end fine-tuning of domain-specific VLMs, leading to biomedical conversational systems with substantial domain knowledge and strong multimodal interaction capabilities (\citealt{li2023llavamedtraininglargelanguageandvision}). In contrast, in solar physics, mainstream general-purpose vision–language models, such as Qwen (\citealt{bai2025qwen25vltechnicalreport}) and Gemini  (\citealt{geminiteam2025geminifamilyhighlycapable}), remain unable to reliably interpret domain-specific scientific imagery (e.g., magnetograms) or perform domain-consistent analytical reasoning, yielding inaccurate or hallucinated responses, as illustrated in Fig.~\ref{fig:general_model}.

In solar physics, the vision-based foundation models through large-scale unsupervised pretraining on solar images, learning generalizable visual representations that can be fine-tuned for downstream tasks such as solar flare prediction (\citealt{2024arXiv241010841R, 2024arXiv241002530W}). However, these scientific foundation models primarily focus on visual feature extraction and lack the language modeling and text generation capabilities required for multimodal reasoning. More recently, Zaman et al. proposed an astronomy-oriented vision–language model based on the LLaVA architecture (\citealt{liu2023visualinstructiontuning}), constructing a diverse image–text dataset of approximately 30,000 pairs to enable natural language interaction with astronomical images. Nevertheless, their dataset includes only limited solar physics imagery, constraining its applicability to solar-specific analysis tasks (\citealt{2025arXiv250408583Z}).

Existing general-domain and astronomy-oriented foundation models exhibit limited capability in interpreting solar magnetograms and related solar imagery. These limitations not only hinder rigorous scientific analysis but also constrain the effective dissemination of domain knowledge in educational and outreach contexts. Motivated by these challenges, we propose JinWu Vision–Language (JW-VL), a fine-tuned foundation model tailored for solar physics. While the current system may not yet meet the rigorous, high-precision demands of operational solar physics, JW-VL establishes a valuable methodological framework for applying multimodal deep learning to this field.  By constructing a multi-task vision–language dataset comprising approximately 60,000 image–text pairs, JW-VL improves multimodal understanding of solar observational images and supports scientific image recognition, analysis, and reasoning through natural language interaction. Building upon JW-VL, we further develop an intelligent agent system for solar activity analysis that automatically generates daily solar activity reports, demonstrating the potential of vision–language models in assisting real-world analytical workflows in solar physics. Specifically, our contributions are summarized as follows:

Building upon JW-VL, we further develop an intelligent agent system for solar activity analysis that automatically generates daily solar activity reports, demonstrating the potential of vision–language models in assisting real-world analytical workflows in solar physics.

1) We introduce JW-VL as a fine-tuned foundation model for solar physics. To our knowledge, it provides the first methodological framework addressing the gap between solar observational imagery and multimodal language-based reasoning, paving the way for advanced deep learning applications in the field.

2) JW-VL supports multiple vision–language tasks, including solar image description, image-based question answering, and optical character recognition (OCR), enabling bilingual (Chinese and English) multimodal dialogue.

3) Based on JW-VL, we develop a “Daily Solar Activity Reports” agent that performs automated analysis of daily solar activity and provides observational recommendations, showcasing how such models can address the real-world requirements of routine space weather monitoring as a valuable assistive tool.


\section{The CoT-SFT Dataset}
\label{sect:data}

This section describes the construction of the Chain-of-Thought Supervised Fine-Tuning (CoT-SFT) dataset, a multimodal solar physics dataset developed to support the training of the JW-VL model. The dataset integrates solar observational data from both ground-based and space-based instruments. Ground-based observations are primarily obtained from a suite of telescopes at the Huairou Solar Observing Station (HSOS), including the 35-cm Solar Magnetic Field Telescope (SMFT; \citealt{Ai1986SolarMagneticTelescope}), the Full-disk Solar Magnetism and Activity Telescope (SMAT; \citealt{2007Solar}), the Solar Full-disk Multi-layer Magnetograph (SFMM; \citealt{tongliyue}), and the full-disk H$\alpha$ telescope (\citealt{2015ASPC..495..395W}). The space-based component incorporates data from multiple missions, including the Solar Upper Transition Region Imager (SUTRI; \citealt{2023RAA....23f5014B}), the Advanced Space-based Solar Observatory (ASO-S; \citealt{gan2019advanced}), the Chinese H$\alpha$ Solar Explorer (CHASE; \citealt{2022SCPMA..6589602L}), and the Geostationary Operational Environmental Satellite (GOES; \citealt{1994BAMS...75..757M}), together with multi-wavelength observations from the Solar Dynamics Observatory (SDO; \citealt{pesnell2012solar}) and the Solar and Heliospheric Observatory (SOHO; \citealt{domingo1995soho}). In addition, the dataset incorporates publicly available space weather observations obtained from online sources. After data integration and curation, the dataset comprises approximately 10,000 high-quality solar images.

To construct the training corpus, we adopt a knowledge distillation framework in which the QVQ-Max model (\citealt{qwen_team_qvq_max_2025}) is used as a teacher model to generate image–text annotations. This process produces image–description pairs, image-based question–answer pairs, and image–OCR pairs. As a result, we obtain a bilingual (Chinese and English) corpus of approximately 60,000 annotated samples, with 
six annotations associated with each image, forming the basis for supervised fine-tuning of JW-VL.

\subsection{Multi-Source and Heterogeneous Data Corpus}

\begin{figure}
   \centering
   \includegraphics[width=\textwidth, angle=0]{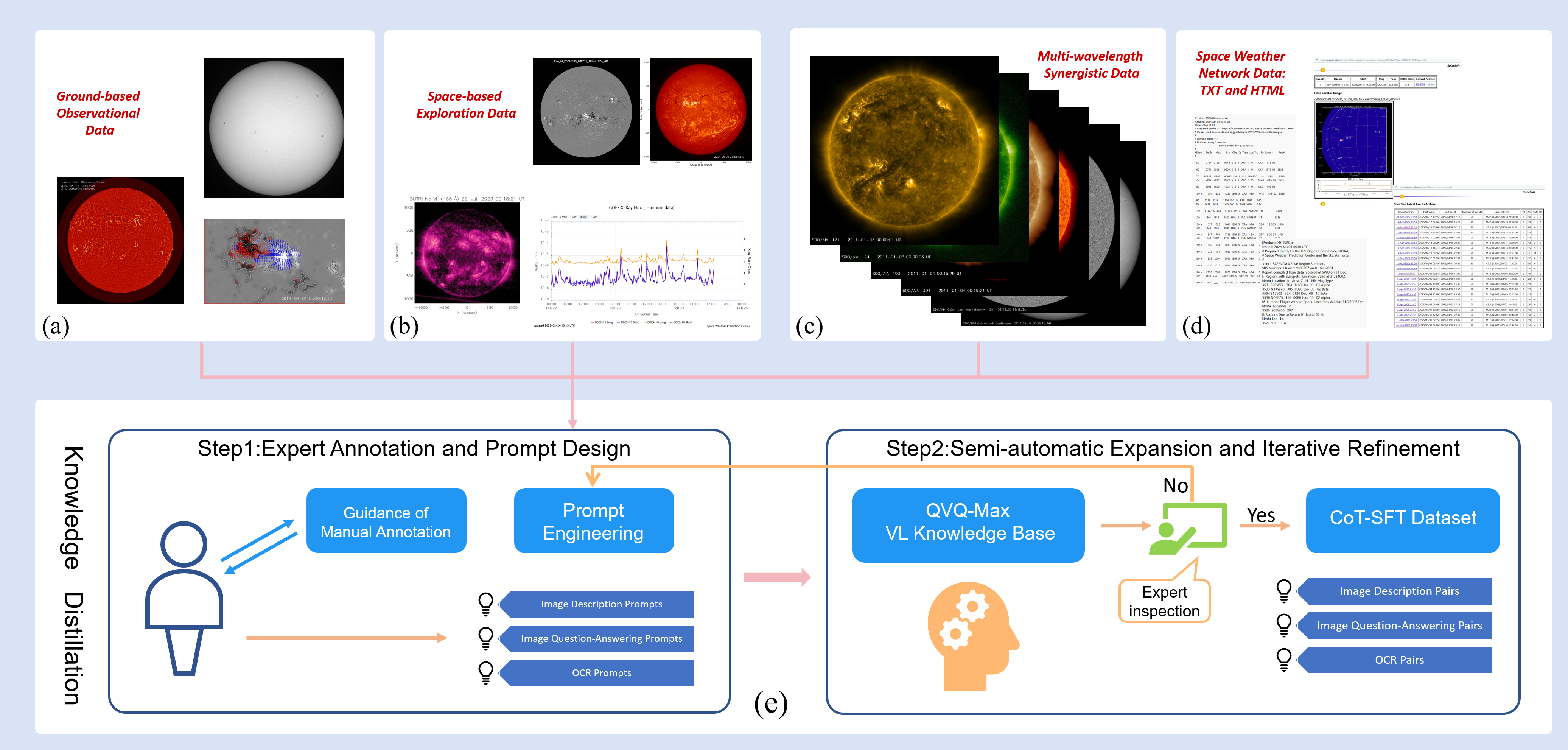}
   \caption{Overview of the pipeline for constructing the CoT-SFT vision--language fine-tuning dataset, including multi-source solar data integration, expert-designed prompt templates, and iterative knowledge distillation.}
   \label{fig:dataset_pipeline}
   \end{figure}

We integrate multi-source and heterogeneous solar observational data to construct a specialized bilingual (Chinese and English) vision–language corpus (\citealt{feng2024roadportabilitycompressingendtoend}). Each sample in the corpus consists of a solar image paired with three types of annotations: (i) a detailed content description, (ii) image-based question–answer pairs, and (iii) OCR results that extract textual information and associated contextual metadata from the image. The data sources are categorized as follows:

1) Ground-based Observational Data: As shown in Fig.~\ref{fig:dataset_pipeline}(a), this category primarily includes observations from ground-based telescopes at HSOS. The data consist of full-disk and active-region photospheric images and magnetograms acquired by the SMFT, SMAT, and SFMM, together with chromospheric H$\alpha$ filtergrams obtained by the HSOS full-disk H$\alpha$ telescope.

2) Space-based Mission Data: As illustrated in Fig.~\ref{fig:dataset_pipeline}(b), this component incorporates observations from several space missions, including extreme-ultraviolet (EUV) images at 46.5~nm from SUTRI, photospheric magnetic field measurements from the Full-disk Vector Magnetograph (FMG) aboard ASO-S (\citealt{2024SoPh..299...70L}), H$\alpha$ observations from CHASE, and soft X-ray flux measurements from GOES.

3) Multi-wavelength Synergistic Data: As depicted in Fig.~\ref{fig:dataset_pipeline}(c), this category incorporates multi-wavelength observations from multiple space-based solar missions, enabling joint analysis across different layers of the solar atmosphere. The data include photospheric continuum images and line-of-sight (LoS) magnetograms acquired by both the Helioseismic and Magnetic Imager (HMI; \citealt{2012SoPh..275..207S}) onboard the SDO and the Michelson Doppler Imager (MDI; \citealt{scherrer1995solar}) onboard the SOHO, providing complementary measurements of the photospheric magnetic field. In addition, multi-channel extreme ultraviolet (EUV) and ultraviolet (UV) observations from the Atmospheric Imaging Assembly (AIA; \citealt{2012SoPh..275...17L}) are utilized, including the 94, 131, 171, 193, 211, 304, and 335~Å EUV channels, as well as the 1600, 1700, and 4500~Å UV and visible continuum channels, jointly capturing plasma structures and dynamics from the photosphere through the chromosphere and into the corona.

4) Space Weather Network Data: As presented in Fig.~\ref{fig:dataset_pipeline}(d), this component comprises structured space weather information retrieved from public online platforms. Key sources include the Solar Region Summary (SRS) and Solar Event Report (ER) provided by the Space Weather Prediction Center (SWPC) of the National Oceanic and Atmospheric Administration (NOAA), as well as monitoring data from the SolarMonitor.org platform.

\subsection{Image-Text Pair Construction}

We design a two-phase knowledge distillation framework to transform the heterogeneous data corpus into aligned image–text multimodal pairs. The framework generates descriptive text, question–answer pairs, and OCR annotations for each image, enabling structured alignment between visual content and linguistic information. The overall pipeline is illustrated in Fig.~\ref{fig:dataset_pipeline}(e).

1) Expert Annotation and Prompt Design: In the first phase, solar physics experts annotate a limited set of images to produce high-quality image–description pairs that serve as seed data. Based on domain expertise, the experts design three categories of prompt templates corresponding to image description, image-based question answering, and OCR. These prompts provide standardized guidance for subsequent automated annotation.

2) Semi-automatic Expansion and Iterative Refinement: In the second phase, the expert-designed prompts are used to guide a knowledge distillation process in which the QVQ-Max model acts as a teacher to generate annotations for a large collection of unlabeled images. The resulting image–text pairs are organized in a structured, stepwise format consistent with a CoT representation. The generated annotations are subsequently reviewed by domain experts, who assign a binary label (“Accept” or “Reject”) based on physical consistency, domain correctness, and the absence of hallucinated content. Accepted samples are directly incorporated into the final vision–language dataset. For rejected samples, prompt templates are revised according to expert feedback, and the distillation process is repeated until satisfactory outputs are obtained. This closed-loop procedure ensures progressive improvement in annotation quality.

The resulting CoT-SFT dataset comprises approximately 60,000 high-quality annotated samples (about 10,000 images), in which each solar image is associated with six complementary annotations, including detailed image descriptions, image-based question–answer pairs, and OCR outputs, generated in both Chinese and English. This dataset provides a reliable data foundation for supervised fine-tuning of the JW-VL model. The iterative construction strategy further enables continuous expansion and refinement as new data and domain knowledge become available.

\section{Method}
\label{sect:method}

   \begin{figure}
   \centering
   \includegraphics[width=\textwidth, angle=0]{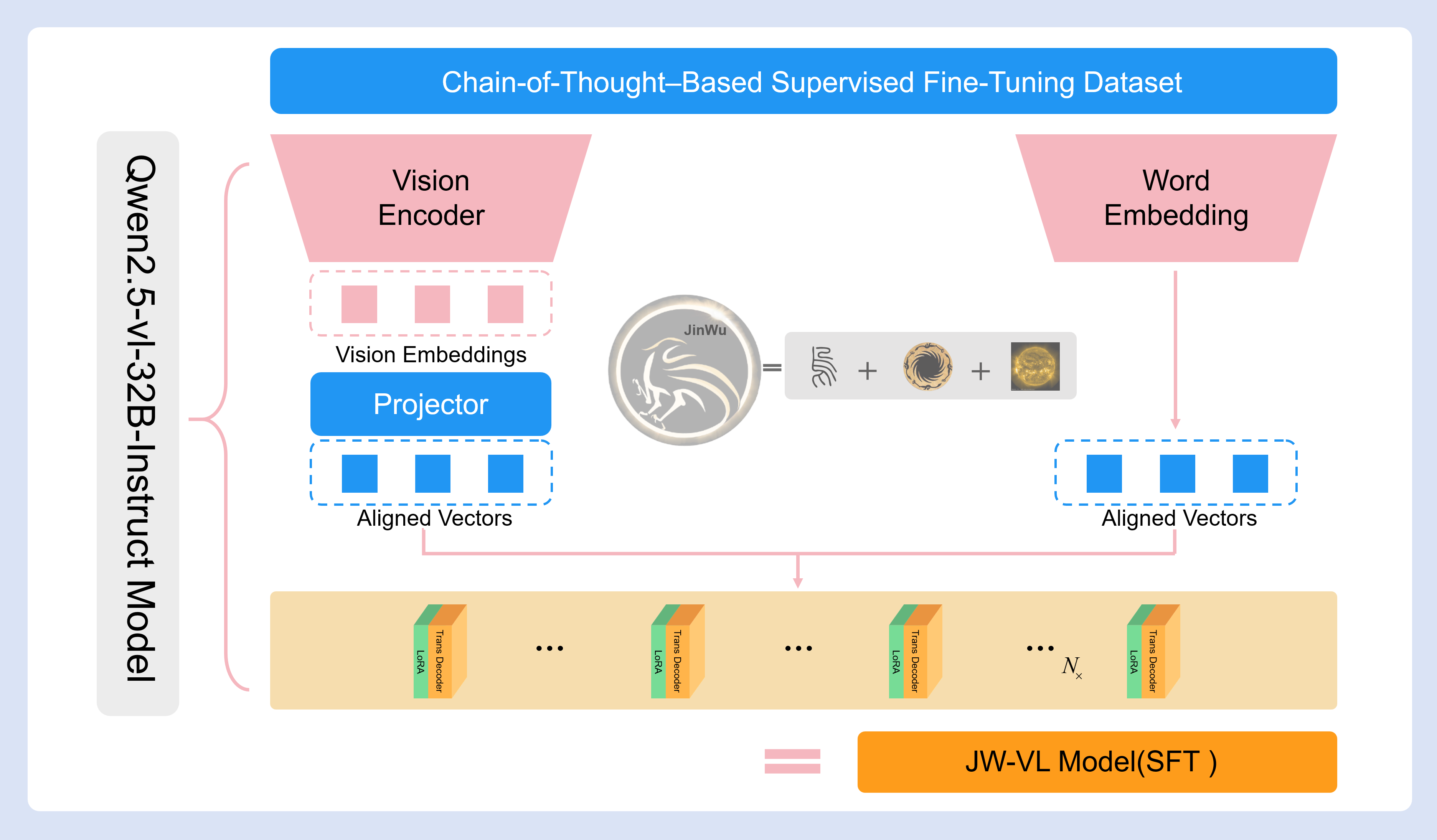}
   \caption{The Architecture of the JW-VL model, illustrating the integration of a vision encoder, cross-modal alignment, and a Transformer-based language decoder fine-tuned for solar physics tasks. The JinWu logo—named for the mythical three-legged bird (JinWu) from ancient Chinese folklore that symbolizes the sun—synthesizes three distinct cultural and scientific elements: The ancient Chinese seal script character: "\raisebox{-0.5ex}{\includegraphics[scale=0.03]{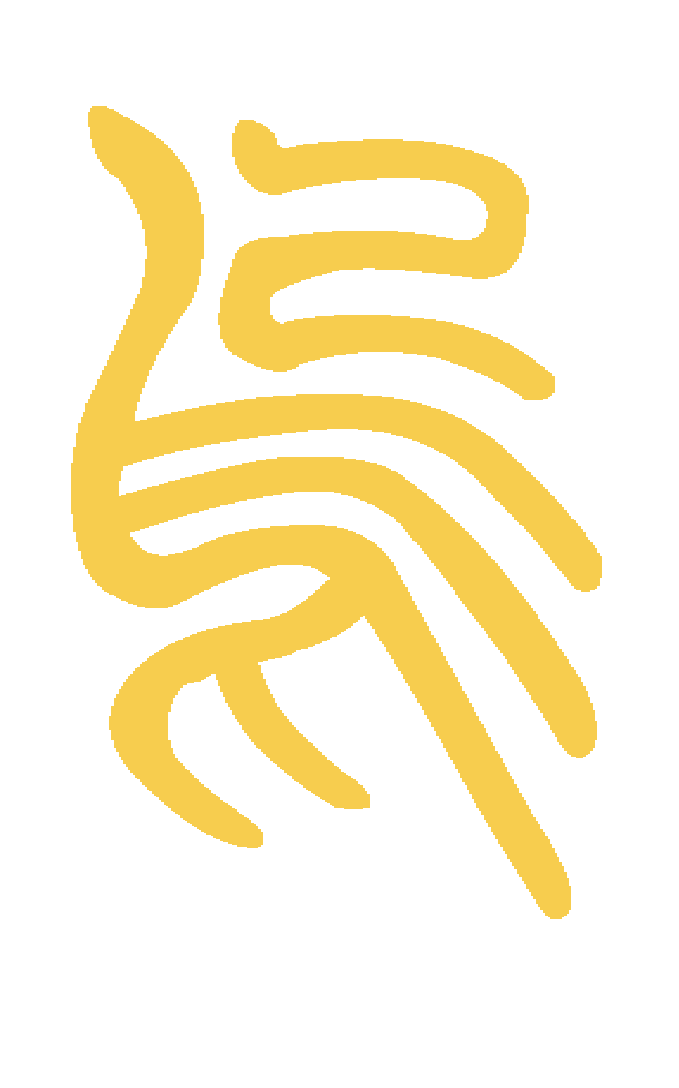}}"; the golden Sun Bird motif -- a culturally significant artifact unearthed at the Jinsha Site (circa 1200-650 BCE) representing solar worship; SDO's Atmospheric Imaging Assembly (AIA) 171 Å band imagery.}
   \label{fig:jwvl_architecture}
   \end{figure}

We constructed JW-VL as a fine-tuned foundation model tailored for solar physics, by performing supervised fine-tuning on the general-purpose Qwen2.5-VL-32B-Instruct model (\cite{2025arXiv250213923B}) using the CoT-SFT dataset developed in this work. JW-VL inherits the general multimodal capabilities of the base model while being explicitly adapted to solar physics imagery, enabling core functionalities including solar image description, image-based question answering and analysis, and OCR.

As illustrated in Figure.~\ref{fig:jwvl_architecture}, Qwen2.5-VL-32B-Instruct is adopted as the base model. It employs a vision encoder capable of fine-grained visual representation learning and supports multimodal inputs involving images and text, providing a suitable foundation for adaptation to scientific visual reasoning tasks. The fine-tuning pipeline proceeds as follows. Textual inputs are first tokenized and mapped into high-dimensional embeddings through a word embedding layer. In parallel, multi-wavelength solar images are divided into visual patches, which are processed by the vision encoder to generate a sequence of visual feature embeddings. These visual features are then projected into the same semantic space as the textual embeddings through a linear projection layer. The resulting visual and textual embeddings are concatenated into a unified multimodal sequence and jointly processed by a multi-layer Transformer decoder to model cross-modal dependencies.

Text generation is performed in an autoregressive manner, with a causal attention mask ensuring that each token prediction depends only on previously generated context. To efficiently adapt the model to solar physics tasks, Low-Rank Adaptation (LoRA) modules are introduced into the attention and feed-forward layers of the Transformer decoder (\citealt{2021arXiv210609685H}). This parameter-efficient fine-tuning strategy substantially reduces the number of trainable parameters and computational overhead, enabling the effective adaptation of a general-purpose foundation model under limited computational resources. At the same time, it preserves the general multimodal capabilities of the base model while facilitating efficient integration of visual and linguistic information. The generation process continues until an end-of-sequence token is produced, yielding coherent text outputs that reflect chain-of-thought reasoning grounded in multimodal inputs.

\section{Experiments and Applications}
\label{sect:agent}

In this section, we evaluate the proposed JW-VL model and demonstrate its applicability to solar physics tasks. We first describe the experimental settings and training configurations adopted for fine-tuning the JW-VL model. We then present both qualitative case studies and quantitative evaluations to comprehensively assess the model’s performance in solar image understanding and multimodal reasoning. To overcome the current lack of publicly available benchmark datasets for vision–language foundation models in solar physics, we curated a dedicated held-out test set to facilitate rigorous quantitative evaluation. Finally, we introduce an agent-based application for daily solar activity reports, illustrating how the proposed model can be integrated into an end-to-end system for automated scientific analysis and report generation.

\subsection{Experimental Setup}

All experiments were conducted on a computing platform running a Linux-based operating system, equipped with four NVIDIA RTX 6000 Ada GPUs. The experimental framework was implemented using PyTorch and ModelScope Swift, a lightweight toolkit designed for efficient fine-tuning of foundation models (\citealt{zhao2024swift}). By leveraging the LoRA technique, the JW-VL model was fine-tuned under low-resource training conditions, enabling efficient adaptation of a 32-billion-parameter vision–language model using a limited number of GPUs.

\subsection{Qualitative Analysis}

We conduct a qualitative evaluation based on a collection of representative case studies, aiming to assess the capability of the JW-VL model in recognizing, analyzing, and interpreting heterogeneous solar observations acquired from both ground-based and space-borne instruments. These cases span multiple wavelengths, observational targets, and task types, reflecting realistic usage scenarios in solar physics research.

   \begin{figure}[t!]
   \centering
   \includegraphics[width=0.85\textwidth, angle=0]{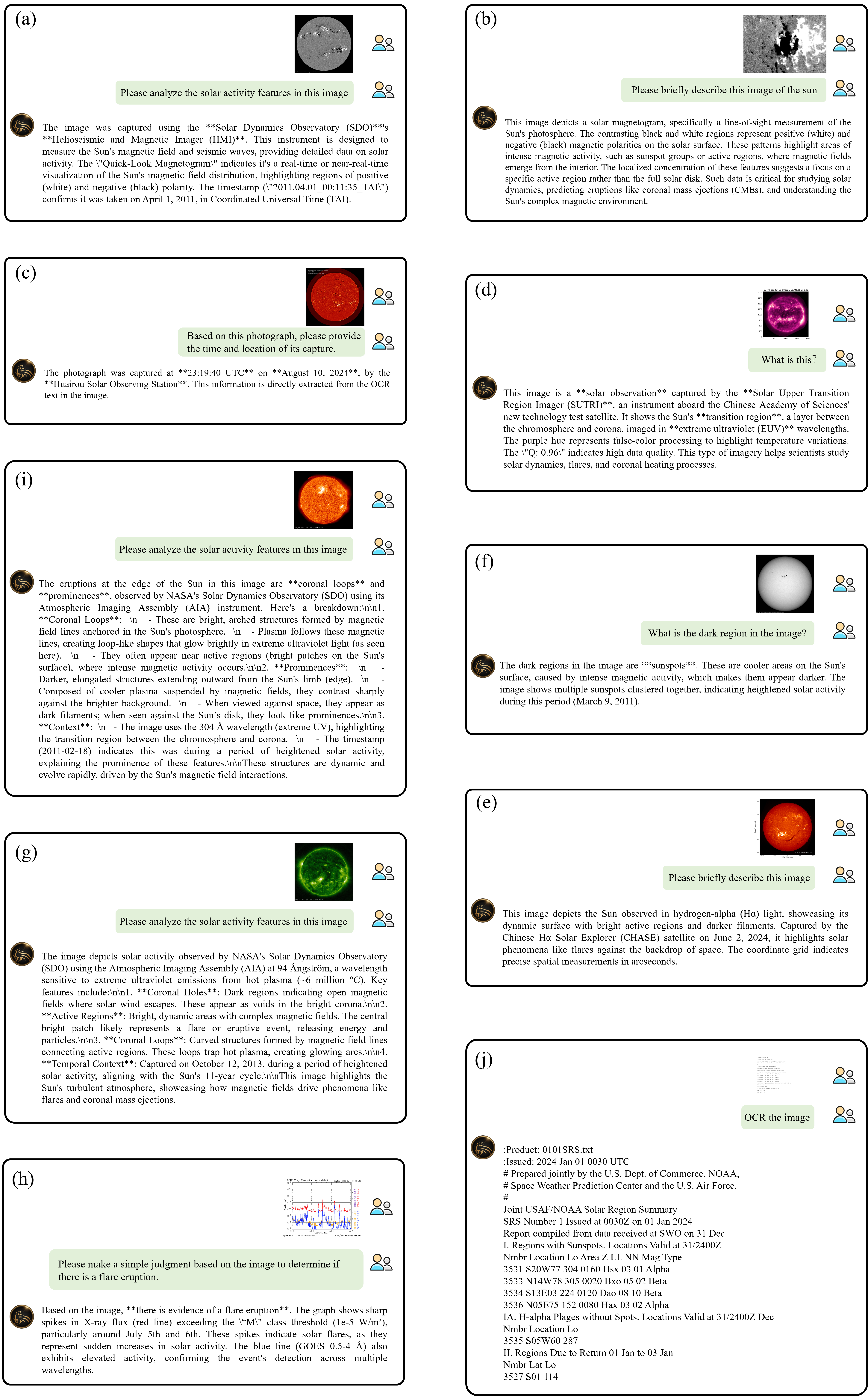}
   \caption{Representative interaction examples illustrating the qualitative performance of the JW-VL model across heterogeneous solar observational data and task types. The examples include recognition and physical interpretation of solar magnetograms and chromospheric images, image-based scientific question answering, identification of solar activity features, and metadata extraction via OCR.}
   \label{fig:magnetogram}
   \end{figure}

The JW-VL model has been deployed in a application and is publicly accessible through the National Astronomical Data Center (NADC) platform \footnote{JW-VL: \url{https://nadc.china-vo.org/ai/chat/chat_index?model_id=3}}. Figure~\ref{fig:magnetogram} presents a series of representative interaction examples illustrating the model’s performance across diverse modalities and vision–language tasks. All evaluation samples were strictly excluded from the training corpus. As shown in Fig.~\ref{fig:magnetogram}(a–c), JW-VL is able to correctly recognize and describe solar magnetograms and chromospheric images, identify localized active regions, and distinguish positive and negative magnetic polarities based on the structural characteristics of the observed magnetic field distributions. The generated explanations are consistent with established solar magnetic field theory.

Beyond localized magnetogram analysis, JW-VL demonstrates physically plausible interpretations for a broad range of solar image types. As illustrated in Fig.~\ref{fig:magnetogram}(d–e), the model supports image-based scientific question answering and physical interpretation across full-disk observations, extreme-ultraviolet imagery, and space-based H$\alpha$ data, enabling the identification of large-scale solar structures, activity-related features, and their potential physical implications.

In addition, JW-VL exhibits the capability to extract and reason over textual and contextual information embedded in solar data products. As shown in Fig.~\ref{fig:magnetogram}(i-j), the model performs metadata-aware OCR and event-related information extraction, allowing it to infer observational time, observing instrument, and associated solar activity records from heterogeneous web-based and observational sources.

Collectively, these examples demonstrate that JW-VL establishes a practical methodological framework, serving as a unified vision–language interface for heterogeneous solar observational data, supporting multimodal understanding, multilingual interaction, and physically consistent reasoning across a wide spectrum of solar physics scenarios. Additional representative interaction examples, including extended multi-image and multi-turn question–answer cases in Chinese, are provided in Appendix A to further illustrate the model’s multilingual reasoning capability and consistency across heterogeneous solar observations.

\subsection{Quantitative Evaluation}

We conducted a quantitative evaluation to assess the effectiveness of the proposed JW-VL model. A held-out evaluation subset (SFT-Test) was constructed using solar images that were strictly excluded from the supervised fine-tuning process. To maintain consistency with the data distribution of the training corpus, the evaluation set follows the same data construction strategy as the SFT dataset.

Specifically, we curated approximately 1000 image interpretation and image-based question–answer pairs covering representative solar observational data from multiple instruments. Reference answers were prepared with the assistance of domain experts in solar physics to ensure scientific correctness and consistency. In addition, for the OCR task, we constructed approximately 200 solar observational images containing embedded metadata. Candidate annotations were initially generated using a high-capability Qwen3.5-Plus model and subsequently verified and corrected by domain experts to establish reliable ground-truth references.

Two representative tasks were evaluated: (1) solar image interpretation and image-based question answering, and (2) OCR-based metadata extraction. To comprehensively evaluate the model's performance, we employed distinct metrics tailored to each task. 

For the image interpretation and question-answering task, we report standard natural language generation metrics, including BLEU-4 and ROUGE-1/ROUGE-2/ROUGE-L, which measure lexical overlap between generated responses and reference answers. For the OCR-based metadata extraction task, strict exact-match accuracy is often overly penalized by minor punctuation shifts. Therefore, we adopted Character-level Accuracy, calculated as $1 - \text{NED}$ (Normalized Edit Distance). Specifically, $\text{NED}$ is defined as $\frac{D(P,G)}{\max(|P|, |G|)}$, where $D(P,G)$ is the Levenshtein (edit) distance between the predicted text $P$ and the ground truth $G$, and $|\cdot|$ denotes the string length. This metric robustly reflects the true capability of the model in recognizing dense, fine-grained solar parameters.

\begin{table}[htbp]
\centering
\caption{Quantitative evaluation results comparing the base model (Qwen2.5-VL-32B-Instruct) and  JW-VL on the SFT-Test dataset. All scores are reported in percentage form.}
\begin{tabular}{l l c c}
\hline
Task Type & Metric & Qwen2.5-VL-32B-Instruct (Base) & JW-VL (Ours) \\
\hline

\multirow{5}{*}{Interpretation \& QA}
& LLM Judge Score & 52.8 & 71.4 \\
& BLEU-4 & 12.68 & 29.69 \\
& ROUGE-1 & 45.10 & 62.82 \\
& ROUGE-2 & 26.61 & 38.82 \\
& ROUGE-L & 35.84 & 48.95 \\

\hline

\multirow{2}{*}{OCR Extraction}
& LLM Judge Score & 53 & 64 \\
& Char-level Acc. & 80.93 & 89.44 \\

\hline

Overall & LLM Judge Score & 52.83 & 70.17 \\

\hline
\end{tabular}
\label{table:Quantitative}
\end{table}

However, automatic text similarity metrics such as BLEU and ROUGE primarily rely on n-gram matching and may not fully capture the scientific correctness or reasoning quality of domain-specific explanations. In scientific contexts, multiple valid descriptions may exist for the same physical phenomenon, and semantically correct explanations may use different terminology or sentence structures. Character-level accuracy only measures the correctness of individual extracted characters and does not evaluate whether the extracted metadata is semantically accurate and consistent with the reference annotations. To address this limitation, we further introduced an LLM-based evaluation framework in which the Qwen3.5 model serves as a judge model. By jointly considering the generated response and the expert-provided reference answer, the judge model evaluates the semantic correctness, physical consistency, and overall quality of the response, producing an aggregated LLM Judge score.

We compared the performance of the base model, Qwen2.5-VL-32B-Instruct, and JW-VL under the same evaluation protocol. Table~\ref{table:Quantitative} demonstrates that JW-VL consistently improves performance across multiple evaluation metrics and provides more physically consistent interpretations of solar observational data.

For the Interpretation \& QA task, JW-VL achieves a substantial improvement in the LLM Judge Score, increasing from 52.8 to 71.4. This gain indicates that JW-VL produces responses that are not only linguistically coherent but also better aligned with the physical meaning of solar observations. Similar improvements are observed in text similarity metrics, where BLEU-4 increases from 12.68 to 29.69, while ROUGE-1, ROUGE-2, and ROUGE-L improve from 45.10 to 62.82, 26.61 to 38.82, and 35.84 to 48.95, respectively. These results suggest that JW-VL generates answers that are significantly closer to the reference annotations in both lexical overlap and semantic content. For the OCR Extraction task, JW-VL also demonstrates improved capability in extracting structured metadata from solar observation images. The LLM Judge Score increases from 53 to 64, while the character-level accuracy improves from 80.93\% to 89.44\%. This indicates that JW-VL not only improves the correctness of individual characters but also enhances the overall semantic consistency of the extracted metadata.

Overall, JW-VL achieves an LLM Judge Score of 70.17 compared to 52.83 for the base model, highlighting the effectiveness of domain-specific training and multimodal alignment for solar physics data interpretation.

\subsection{The Daily Solar Activity Reports Agent}

\begin{figure}
   \centering
   \includegraphics[width=\textwidth, angle=0]{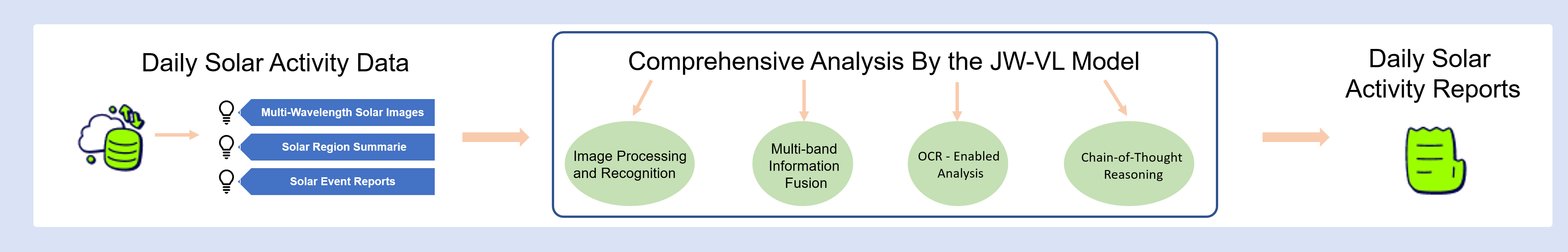}
   \caption{The workflow of the daily solar activity reports agent used as an application example of the JW-VL model. The agent performs automated data acquisition, multimodal interpretation of solar observations, and daily report generation.}
   \label{fig:Agent}
   \end{figure}

We developed a Daily Solar Activity Reports (DSAR) agent as an  demonstration of the practical utility of the JW-VL model in routine solar activity monitoring. The overall architecture of the system is illustrated in Figure~\ref{fig:Agent}. The agent is designed to automate the workflow from heterogeneous data acquisition to multimodal analysis and structured report generation, providing qualitative assessments intended to assist expert interpretation in solar physics and space weather operations.

The system adopts a layered architectural design to enhance modularity and extensibility. The Data Acquisition Layer is responsible for the automated retrieval of heterogeneous solar activity data, including multi-wavelength solar images, SRS, and Solar Event Reports, from authoritative public sources such as SolarMonitor.org and the SWPC/NOAA. To ensure temporal consistency and data integrity, this layer incorporates structured parsing, dynamic content handling, and schema-based validation mechanisms.

The Intelligent Analysis Layer constitutes the core of the agent and integrates the JW-VL model. Through task-oriented prompt templates, JW-VL performs joint interpretation of visual observations and associated textual records, leveraging its multimodal reasoning and OCR capabilities. Based on these analyses, the agent automatically generates structured daily reports that summarize the overall solar activity level, characterize individual active regions, analyze magnetic field complexity, provide qualitative assessments of potential space weather impacts, and highlight regions requiring further attention. The User Interaction Layer provides a web-based interface that enables users to conveniently access the generated reports and inspect the underlying multimodal analysis results.

An illustrative example of the agent’s behavior is provided by the analysis of NOAA Active Region 14294, as shown in Appendix B. Shortly after its emergence, the DSAR agent identified this region as a high-risk active region based on a combination of rapidly increasing area (from approximately 0160 to 0800 millionths of the solar hemisphere) and a pronounced transition in McIntosh classification from Dso to Fkc. This evolution indicates rapid sunspot group development, increasing magnetic complexity, and strong shear, which are commonly regarded as empirical precursors of major flaring activity or coronal mass ejections (CMEs). Subsequent observations over the following two weeks confirmed frequent flare activity associated with this region, including multiple C-class and M-class events, qualitatively consistent with the agent’s early risk assessment.

It is important to note that the DSAR agent should be regarded purely as an assistive analysis tool for qualitative interpretation rather than an operational flare forecasting system. While the agent successfully identified heuristic risks (e.g., complex magnetic configurations) in the specific example presented in this paper, this represents a special, prominent case rather than a generalized predictive capability. 
Instead, it serves as an assistive tool that synthesizes heterogeneous observations and textual records into coherent, human-readable reports, supporting situational awareness and routine analysis.

Overall, this agent demonstrates the feasibility of integrating a  fine-tuned foundation model into an end-to-end automated analysis pipeline for solar activity data. While the present implementation focuses on qualitative interpretation and report generation, future work will be integrated with our dedicated JW-Flare model to transition its flare assessment capabilities from qualitative interpretations to accurate, data-driven quantitative forecasting (\cite{shao2025jwflareaccuratesolarflare}).

\section{Conclusions}
\label{sect:conclusion}

In this study, we presented JW-VL, a fine-tuned foundation model tailored for solar physics, with the aim of addressing the limitations of general-purpose vision--language models in interpreting specialized solar observational data. Through supervised fine-tuning on a large-scale, multi-source, and bilingual solar physics dataset, JW-VL acquires domain-aware multimodal capabilities, including solar image description, image-based question answering, and OCR, while supporting both Chinese and English interactions. Although the current architecture serves primarily as an assistive tool for qualitative interpretation rather than a fully operational, high-precision forecasting instrument, this research successfully lays a methodological groundwork for adapting multimodal deep learning to solar physics.These capabilities enable more effective understanding of heterogeneous solar observations and associated textual records.

Building upon the JW-VL model, we further developed an intelligent agent for daily solar activity analysis and reporting. By integrating automated data acquisition, multimodal interpretation, and structured report generation, the agent demonstrates the practical applicability of JW-VL in real-world scenarios. The system supports routine analysis of solar activity, identification and characterization of key active regions, qualitative assessment of magnetic field complexity, and generation of coherent daily summary reports, thereby providing a practical tool for solar physics research and space weather monitoring.

Recent state-of-the-art models, such as Gemini 3 Pro, have shown encouraging progress in recognizing certain solar observational products, including localized LoS magnetograms; however, consistent and accurate interpretation of the full range of domain-specific solar data remains challenging. While future general-purpose models may eventually incorporate much of this specialized knowledge, such advances are unlikely without dedicated datasets, task formulations, and problem settings contributed by domain-focused studies. In this regard, JW-VL provides domain-specific data resources and structured evaluation scenarios that help bridge this gap. It is designed to build upon the evolving multimodal reasoning capabilities of general foundation models, enabling the development of intelligent assistants that support routine scientific analysis. Progress in general-purpose models is therefore complementary to, and directly beneficial for, the long-term development of such research-oriented systems.

Guided by the real-world requirements of the solar physics region, our future work will proceed along several critical dimensions. First, recognizing that operational astrophysics relies heavily on raw, scientific-grade data, we will develop a comprehensive processing pipeline to directly ingest and interpret FITS (Flexible Image Transport System) files. This approach will transcend current limitations associated with pseudo-color images, enabling accurate analysis of intrinsic structural features in original multi-wavelength grayscale observations without relying exclusively on fine-tuning. Second, we aim to evolve the architecture of JW-VL towards a seamlessly fused vision–language paradigm (akin to state-of-the-art models like Qwen3.5-VL), eliminating the explicit boundaries between visual and textual modules to maximize its reasoning capabilities in practical applications. Third, at the functional level, we will construct a robust, structured solar physics knowledge base to elevate the model’s analytical outputs to strict professional standards. Finally, by integrating advanced computer vision capabilities—such as precise object detection and segmentation for key solar phenomena—we intend to significantly broaden the model's utility for researchers. Ultimately, our goal is to evolve JW-VL into a comprehensive, intelligent research assistant that is fundamentally driven by the practical applications of solar physics.

\begin{acknowledgements}

 This research was supported by the National Astronomical Observatories, Chinese Academy of Sciences (No. E4TQ2101), and the National Natural Science Foundation of China (No. 12473052), with additional support from the Specialized Research Fund for the State Key Laboratory of Solar Activity and Space Weather. Part of this work was presented as an oral report at the Third Workshop on Astronomical Big Data and Artificial Intelligence, where it received the Outstanding Student Presentation Award. The award was accompanied by financial support provided by the editorial office of Research in Astronomy and Astrophysics (RAA). Furthermore, we extend our appreciation to the NADC for their technical support in deploying the JW-VL model.

\end{acknowledgements}


\bibliographystyle{raa} 
\bibliography{bibtex}

\appendix                  

\section{Additional Qualitative Interaction Examples:}

This appendix presents additional qualitative interaction examples of the JW-VL model, complementing the experimental results discussed in Section 4.2. The examples focus on multi-image and multi-turn vision–language interactions conducted in Chinese, covering solar image recognition, physical interpretation, activity assessment, and metadata extraction. They are intended to further demonstrate the robustness and multilingual applicability of JW-VL under realistic solar physics usage scenarios.

   \begin{figure}
   \centering
   \includegraphics[width=\textwidth, angle=0]{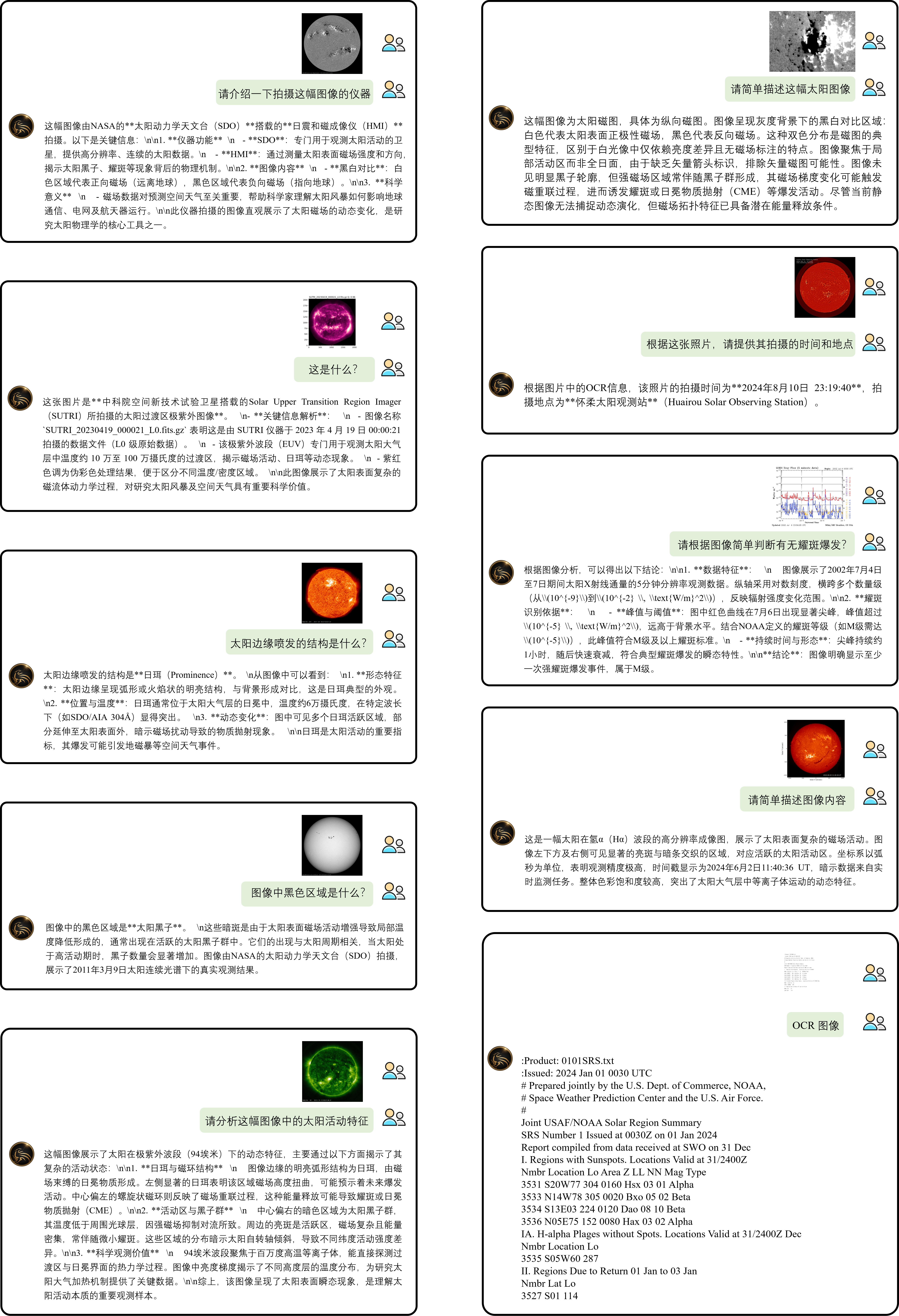}
   \caption{Additional qualitative examples of vision–language interactions of JW-VL in Chinese.}
   \label{fig:results-CN}
   \end{figure}

\section{The daily solar activity reports:}

This appendix presents a representative example of a daily solar activity report automatically generated by the proposed analysis agent, as discussed in Section 4.3. The report illustrates the structured output produced by the agent through heterogeneous solar observations and associated textual records, including multi-wavelength imagery, SRS, and event-related metadata. The presented example demonstrates how the agent synthesizes observational information into a coherent, human-readable daily report, covering overall solar activity assessment, characterization of individual active regions, qualitative analysis of magnetic field complexity, and preliminary discussion of potential space weather implications. The example is intended to illustrate the format, content organization, and qualitative reasoning capability of the agent.

   \begin{figure}
   \centering
   \includegraphics[width=\textwidth, angle=0]{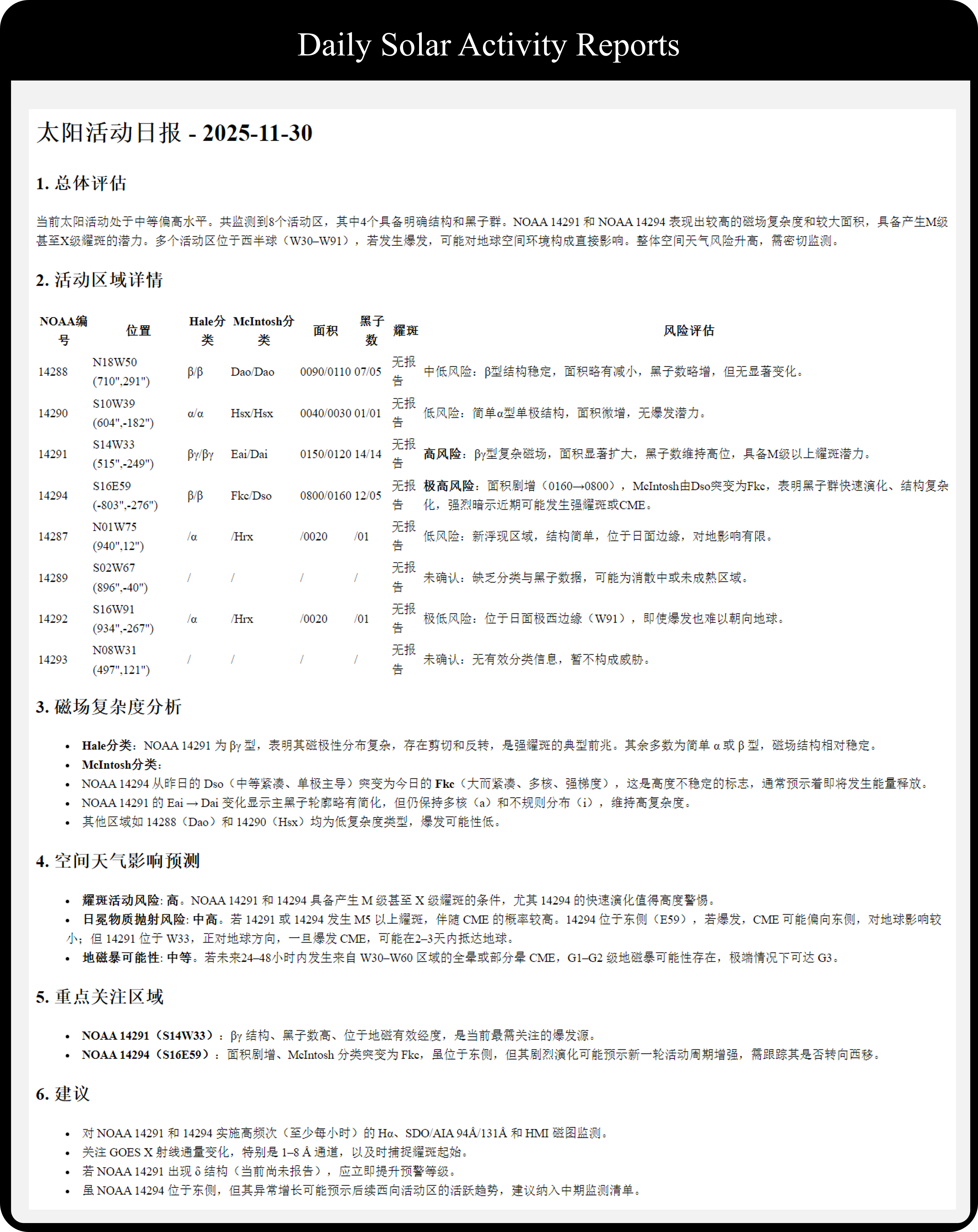}
   \caption{Example of an automatically generated daily solar activity report produced by the DSAR agent.}

   \label{fig:report}
   \end{figure}

\label{lastpage}

\end{document}